\documentclass[proof]{WileyASNA-v1}

\articletype{Article Type}%

\received{28 August 2018}
\revised{10 October 2018}
\accepted{10 November 2018}

\begin{document}

\title{Automatic detection of tidal disruption events and other long duration transients in XMM-Newton data}

\author[1]{Natalie A. Webb}



\authormark{Webb}

\address[1]{\orgdiv{IRAP}, \orgname{Universit\'e de Toulouse, CNRS, CNES}, \orgaddress{\state{Toulouse}, \country{France}}}



\corres{*Corresponding author \email{Natalie.Webb@irap.omp.eu}}


\abstract{XMM-Newton's large field of view and excellent sensitivity have resulted in hundreds of thousands of serendipitous X-ray detections. Whilst their spectra have been widely exploited, their variable nature has been little studied. Part of this is due to the way XMM-Newton currently operates, where observations generally have a 12 month proprietary period. It is often too late to follow-up a serendipitous transient a year after detection. New robust software could be introduced into the pipeline to automatically identify bright transients that are not the target of the observation. Statistically, hundreds of tidal disruption events (TDEs) have been detected serendipitously by XMM-Newton. With prior consent from the PI of the observation, an automatic alert to a new transient could be set up, allowing it to be followed-up within weeks, ideal for TDEs that are bright for about a year. Over the next decade, hundreds more TDEs should be detected. Following-up the brightest in quasi-real time would allow constraints to be made on the black hole mass, spin and accretion regime and identify intermediate-mass black holes that are expected to be hidden in faint, low-mass galaxies. This article discusses the advantages that such changes would have on the follow-up of transients and TDEs.}

\keywords{X-rays: general, catalogs, accretion, accretion disks, Galaxy: evolution}

\fundingInfo{CNES}

\maketitle


\section{Introduction}\label{sec:intro}

Two varieties of black hole candidates (BHs) have been widely observed to date: stellar mass ($\sim$3-100 M$_\odot$) black holes \citep[e.g.][]{bolt72} and supermassive ($\sim$10$^{6-10}$ M$_\odot$) black holes \citep[e.g.][]{lynd69}, present in the cores of massive galaxies. It is believed that stellar mass black holes are formed from the collapse of massive stars or the mergers of two neutron stars, but  how supermassive black holes (SMBH) are formed and evolve is unclear. They can not form from stellar mass black holes, as even continuously accreting at the Eddington limit, they would never reach masses as high as $\sim$10$^9$  M$_\odot$ observed in a massive quasar at z$\sim$7.1 \citep{mort11} or the 8$\times$10$^8$ M$_\odot$ black hole found at z=7.54 \citep[0.69 Gyr,][]{bana18}. Different theories propose that smaller, intermediate mass black holes (IMBH, 10$^{2-5}$ M$_\odot$) would either merge and/or accrete to create SMBH \citep[see][for reviews]{volo12,gree12,mezc17}. This may be at or above the Eddington rate, although the physical mechanism for super-Eddington accretion is still to be elucidated.   If this is the case, we expect to detect IMBHs, as many of them will avoid such mergers or may even be ejected during merger interactions.  Indeed, \cite{mada01} suggest that there may be as many as 10$^3$-10$^4$ such black holes in some galaxies. However, until recently, the observational evidence for  IMBHs was very weak, making it difficult to study how they accrete matter and validate the theory on how SMBH form. 

IMBH are thought to originate either from the implosion of massive population III stars, the first stars to be formed in the Universe, creating IMBH of $\sim$100 M$_\odot$, or from the direct collapse of low metallicity dust clouds, again in the early Universe, creating IMBH of $\sim$1000 M$_\odot$ \citep{mill04}. Their presence in the early Universe is thought to have contributed significantly to the ultra-violet  background, as the IMBHs would have shone brightly as they accreted matter \citep{kawa03}, implying that IMBHs may have participated in the cosmological re-ionisation \citep{mada04}.  Finding relic IMBHs or measuring the masses of the black holes at the lower end of the massive black holes range will allow us to understand the mass of the black hole seed. For example, \cite{gree12} and references therein show that if the IMBHs originate from $\sim$1000 M$_\odot$ black holes created for example through the collapse of low metallicity dust clouds in the early Universe, by the present day, those that have avoided merging with other IMBHs would be found in low mass galaxies of $\sim$10$^9$ M$_\odot$ and about half would have had the opportunity to accrete matter and to reach masses of $>$ 10$^5$ M$_\odot$ today. If the IMBHs form from less massive seeds of $\sim$100 M$_\odot$, those that have avoided merger would again be in the low mass galaxies $\sim$10$^9$ M$_\odot$, but in this case $\sim$90\% of the massive black holes in their centres would have a mass of $\sim$10$^4$ M$_\odot$. 

It is however difficult to find IMBH, as they are often accreting at a very low level and/or are located at large distances. Traditionally they have been searched for in the centres of low mass galaxies, as the mass of the central black hole has been shown to scale with the galaxy mass \citep[][]{ferr00}, or in the centres of stellar clusters \citep[e.g.][]{hut92}. Alternatively, they maybe found in the outskirts of galaxies or clusters as they merge with the central SMBH, or they may be the compact object in the brightest ($>10^{41}$ erg s$^{-1}$) ultra luminous X-ray sources (ULXs) \cite{mezc17}.

Hyper Luminous X-ray source 1 (HLX-1) is an example of an extreme ultra luminous X-ray source which has a maximum X-ray luminosity of 1.2 $\times$ 10$^{41}$ erg s$^{-1}$ \citep{gode09}. It was discovered in the {\it XMM-Newton} serendipitous source catalogue whilst searching for new Galactic and extra-galactic stellar mass compact objects \citep{farr09}. The subsequent validation of HLX-1 as an intermediate mass black hole \citep[][and references therein]{webb12,gode14,yan15,webb17} was the first step in identifying the population of IMBH that we expect to detect. We can use what we have learnt from this object to search out and validate other IMBH. One of the major difficulties in detecting IMBH using electromagnetic observations is that they are only bright when they are accreting significant amounts of material, which is difficult to achieve \citep[e.g.][]{mill04}. Indeed, HLX-1 is only periodically bright, when sufficient material is available for accretion, probably through tidal stripping of a companion star at periastron \citep{laso11,gode14}. For a system to be in such a configuration is rare, which could explain why similar systems have not been detected in large numbers. Indeed, similar objects have been looked for and although some good candidates have been identified, they are not yet confirmed as bonified IMBH \citep[e.g.][]{zolo16,sutt12}.

\begin{figure}[t]
\centerline{\includegraphics[width=250pt,height=18pc]{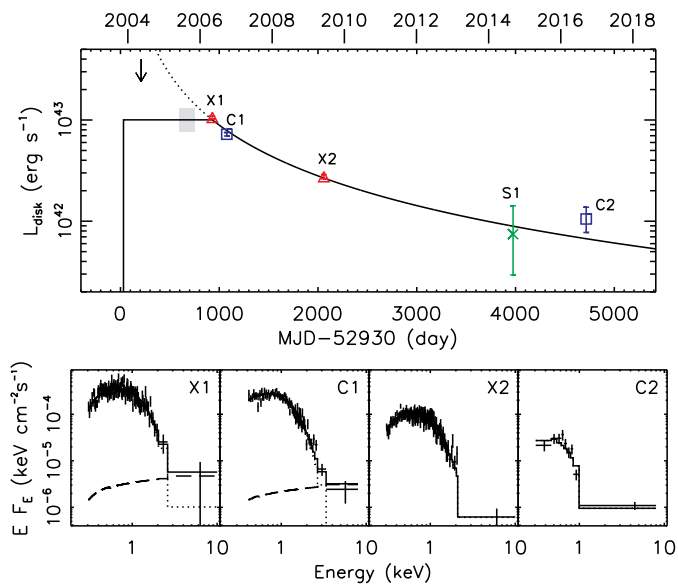}}
\caption{The long-term luminosity and spectral evolution of the TDE J2150-0551. Upper panel: the bolometric disk luminosity curve. The Chandra (C1, C2), XMM-Newton (X1, X2) and Swift (S1) pointed observations are shown as blue squares, red triangles and green cross, respectively, with 90\% error bars, and the arrow shows the 3$\sigma$ upper limit from the XMM-Newton slew observation on 14 May 2004. The gray shaded region indicates when an optical flare was detected in 2005. The solid line shows the typical exponential decay observed in TDEs.  Lower panel : Spectra of the XMM-Newton and Chandra observations showing the characteristic very soft thermal emission. Figure reproduced with permission from \cite{lin18}.\label{fig:TDEdacheng}}
\end{figure}

HLX-1 is exceptional in as much as it appears that the companion star has avoided total tidal disruption. A star approaching the black hole slightly closer would be ripped apart (a tidal disruption event, TDE), with approximately half of the matter falling on to the massive black hole \cite{rees88} and causing the system to become brighter by several decades in luminosity in X-rays  and at other wavelengths before decaying back to the original luminosity over years, e.g. \cite{holo14,blag17,cenk12,lin11}, and see also the Open TDE database\footnote{https://tde.space/}.  A typical TDE X-ray lightcurve can be seen in Fig.~\ref{fig:TDEdacheng}, upper panel, with its characteristic exponential decay in luminosity proportional to $\sim$time$^{-5/3}$ \citep[e.g.][]{evan89},  reflecting the  timescale  on  which  the  stellar  debris  eventually  returns to pericentre, shown with a solid line. 

TDEs are intrinsically interesting as accretion is expected to exceed the Eddington limit at the outset \citep{evan89}. Studying the emission may provide clues to the physical mechanism behind this phenomenon, helping us to understand the growth of SMBH. Observing TDEs could also help us understand why the majority show soft X-ray spectra (see for example the lower panels in Fig.~\ref{fig:TDEdacheng}), but a few show hard X-ray emission. This may be due to a jet pointing towards us \citep[e.g.][]{auch17}, but other mechanisms are also suggested \citep{hryn16}. Further observations may also help us to understand if the viewing angle causes some TDEs to be seen in the optical domain and not in the X-ray e.g. PTF-09ge, whilst others in the X-ray and not the optical, and others still are seen in both \citep{auch17}, as suggested by \cite{dai18}. To further complicate things, very recently, a TDE event was observed in the radio domain, but not in optical or X-rays \citep{matt18}.

However, as the tidal radius of the black hole must be outside of the Schwarzschild radius for us to observe the tidal disruption event, TDEs with main sequence stars are generally only detected for the lower mass black holes \citep[$<$ 10$^8$ M$_\odot$][]{rees88}. A reasonable proportion of the detected TDEs will be IMBH, as the number density of lower mass galaxies, which have a tendency to house the lower mass black holes, dominates over that of the more massive galaxies \citep[see e.g.][]{torr15}. Therefore searching for TDEs is a good way to identify new IMBH and detecting them with X-ray observations, it is possible to determine the mass of the black hole by fitting the soft thermal emission with disc models, e.g. \cite{lin18,lin15,gode12}. If the event is close enough and therefore bright enough to detect an iron line in the spectrum, this would also allow us to put limits on the mass and the spin (e.g. Karas et al. 2014).

\section{Tidal disruption events in XMM-Newton data}\label{sec:TDEsXMM}

\begin{figure}[t]
\centerline{\includegraphics[width=320pt]{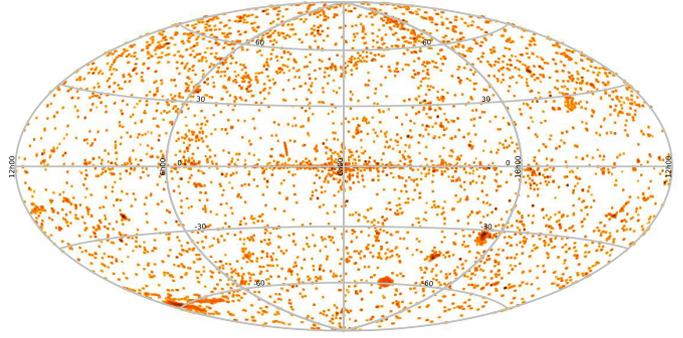}}
\caption{The detections in the 3XMM-DR8 catalogue shown in a Hammer-Aitoff projection. The darker the colour, the greater the number of observations.\label{fig:3XMMcat}}
\end{figure}

Many TDEs have been identified through exploring {\it XMM-Newton} data, e.g. \cite{lin11,saxt15,lin17a,saxt17,lin18} and notably through exploring the {\it XMM-Newton} catalogue \citep{rose16}, produced by the {\it XMM-Newton Survey Science Centre}\footnote{http://xmmssc.irap.omp.eu/} (SSC) \citep{wats01}. The most recent version of this catalogue is 3XMM-DR8 and was released in May 2018\footnote{http://xmmssc.irap.omp.eu/Catalogue/3XMM-DR8/3XMM\_DR8.html}. It contains 775153 X-ray detections, where objects have been detected as many as 59 times over the 17 years from February 2000 to November 2017. 332 columns of information are provided for each detection, including coordinates, observation date, time and mode, exposure and background information, counts, fluxes and rates in 7 energy bands, maximum likelihoods of detection, quality and variability flags, as well as multi-band images, lightcurves and spectra. The distribution of X-ray detections in 3XMM-DR8 on the sky can be seen in Fig.~\ref{fig:3XMMcat}. More recently still, the SSC has released the first stacked catalogue of sources, 3XMM-DR7s\footnote{http://xmmssc.irap.omp.eu/Catalogue/3XMM-DR7s/3XMM\_DR7stack.html} \citep{trau18}, where for each source identified from the stacked detections, information regarding each detection is provided, along with a long term lightcurve, allowing easy access to the long term variability of sources.

Several TDEs have been detected in the {\it XMM-Newton} slew survey \citep[e.g.][]{saxt08,saxt12,saxt14}. This is thanks to a systematic comparison of slew survey detections as they are found, with the same region of the sky observed by previous missions. The comparison is made with the {\it Rosat All Sky Survey} \citep[RASS][]{voge99,voge00,boll16}, which has a similar depth and position error to the slew survey \citep{saxt08}, as well as the {\it Einstein} slew survey, the {\it Exosat} slew survey, the {\it HEAO-1/A2} and {\it RXTE} slew surveys, which have shallower flux limits \citep{saxt08}. This approach is very successful as it allows the follow-up of the TDE rapidly whilst the source is still bright and not a year or years later, when the source is examined as a part of the catalogue. TDEs have none the less been detected using the {\it XMM-Newton} catalogue e.g. \cite{khab14,lin15,feng15,lin17b}, even though no systematic comparison has been carried out at the time of the detection. Further, limited information can be extracted from the observations as no follow-up was done within a year of the detection, due to the data proprietary time of 12 months.  

The median flux in the {\it XMM-Newton} soft band (0.2-2.0 keV) for the latest catalogue release, {\it 3XMM-DR8}, is 5.2$\times$10$^{-15}$ erg cm$^{-2}$ s$^{-1}$. These pointed observations are much deeper than the typical flux limit of the {\it RASS} of 3$\times$10$^{-13}$ erg cm$^{-2}$ s$^{-1}$ (0.2-2.4 keV) \citep{voge00}. Only 21136 sources in {\it 3XMM-DR8} ($<$3\%) have 0.2-2.0 keV fluxes greater than the {\it RASS} flux limit, thus limiting the number of new TDE identifications by comparing the {\it XMM-Newton} flux to similar regions of the sky observed with {\it Rosat}, as for the {\it XMM-Newton} slew survey. However $\sim$100000 {\it 3XMM-DR8} sources have been observed at least twice (and up to 59 times) and further sources have been observed with {\it Chandra}, which reaches similar flux limits to {\it XMM-Newton}, over a baseline of 18 years. This is the equivalent of $\sim$1300 fields, based on the number of detections in {\it 3XMM-DR8} and the number of fields observed. Assuming that the observations were in full frame mode (where 79\% of MOS 2 observations are in full frame mode), means that $\sim$260$^\circ$ squared have been surveyed at least twice with {\it XMM-Newton}, and more if we consider fields observed with {\it XMM-Newton} and {\it Chandra}.

In an average length XMM-Newton observation of ~37 ks (average value from {\it 3XMM-DR8}) we can detect a TDE out to z$\sim$1.5 (for a luminosity of 10$^{43}$ erg s$^{-1}$, typical of TDEs at the Eddington luminosity for a black hole of 10$^5$ M$_\odot$) and supposing that the absorption towards the galaxy is not excessively high. In the {\it XMM-Newton} 30' diameter field of view, there are about 10$^5$ galaxies out to z$\sim$1.5, based on the Hubble Deep Field observations and the number of galaxies per square degree per redshift, e.g. \cite{sand05}. However, given that the TDE rate is 1.7$\pm$$^{\scriptscriptstyle +2.85}_{\scriptscriptstyle -1.27}$ $\times$ 10$^{-4}$ gal$^{-1}$ yr$^{-1}$ \citep[90\% confidence,][]{hung18}, in an average length {\it XMM-Newton} observation, there is then a probability of $\sim$0.05 that a detectable TDE is occurring in a field. As TDEs remain bright for about a year, this increases the number. Taking into account 71\% of {\it 3XMM-DR8} observations are above and below the Galactic plane ($\pm$15$^\circ$, see also Fig.~\ref{fig:3XMMcat}), where most TDEs will be observed, around 700 TDEs should exist in current {\it XMM-Newton} data. However, to identify  them we require another 'deep' X-ray observation for comparison. Statistically, this will available for the $\sim$1300 fields, as described above, so around 100 TDEs should be detectable using {\it XMM-Newton} data only. Using {\it Chandra} data and other X-ray observations will increase the number that can be identified. An observation has on average 76 X-ray detections (calculated from the number of observations and detections in {\it 3XMM-DR8}) and TDEs are variable and often show low energy thermal emission in X-rays, so it is much easier to identify the TDE in X-rays than in the optical.

Technically, it is possible for {\it XMM-Newton} to continue observing for another 10 years or so (until 2029/2030). At least 50 new TDEs should be detectable in this time using only {\it XMM-Newton} as a baseline. However, as of 2019, {\it eROSITA} \citep{pred11} will fly and the all sky survey will have a flux limit of $\sim$1$\times$10$^{-14}$ erg cm$^{-2}$ s$^{-1}$ \citep[0.5-2.0 keV][]{pred14}, close to the median {\it XMM-Newton} observations, and thus providing a point of comparison for most {\it XMM-Newton} pointings and allowing around 200 new TDEs to be identified. If these new TDEs can be followed up rapidly, this would help us identify new IMBH, see Sec.~\ref{sec:intro}.

\section{Discussion}\label{sec:disc}

So how would we go about following up TDEs (and other transients) rapidly? Firstly, during the pipeline processing of the {\it XMM-Newton} data, the flux of each source should be automatically compared to previous X-rays observations of the same region of sky. Sources showing extreme flux variations compared to previous pointings, would be noted. The coincidence with the centre of a known galaxy would point to a TDE, although changing look AGN can also show significant variability, but this is invariably a factor of a few \citep[e.g.][]{ricc16,zetz18,hern15} and not the factor of hundreds or thousands that is often associated with TDEs, e.g. \cite{lin18b,saxt12,auch17}. Some ULXs exhibiting luminosity variability of a factor 1000 are also found towards the centres of galaxies \cite{walt15}, but it has been shown that the central compact object is a neutron star and the maximum X-ray luminosity does not generally exceed $10^{41}$ erg s$^{-1}$ \citep[e.g.][]{bach14,isra17,fuer16}, so these should be fairly easy to distinguish from TDEs.

With prior consent from the PI of the observation, similar to when PIs gave prior consent to the SSC to follow-up serendipitously detected X-ray sources that were not the target of the observation (at the beginning of the mission), an automatically generated message could alert the PI, or even the community at large, to the new transient, allowing it to be followed-up within weeks, ideal for TDEs that are usually bright for about a year. 

Obviously other highly significant transient events could be identified in a similar way, such as the electromagnetic counterparts to gravitational wave events, $\gamma$-ray burst (afterglows), cataclysmic variable outbursts, supernovae, X-ray binary outbursts, magnetar bursts, etc. Whilst the follow-up of a magnetar would not be feasible given their short outburst duration, this method would allow them to be identified automatically. The transient nature of the other objects is usually weeks to years, allowing them to be followed. This would add a new dimension to the use of {\it XMM-Newton}, which can detect much fainter objects than for instance {\it MAXI} which reaches $\sim$7$\times$10$^{-10}$ erg cm$^{-2}$ s$^{-1}$ (2-30 keV) per orbit \citep{mats09} or the {\it Neil Gehrels Swift Observatory} ({\it Swift}) which reaches 1$\times$10$^{-13}$ erg cm$^{-2}$ s$^{-1}$ for blind searches in a 10 ks observation (0.2-10 keV)\footnote{https://swift.gsfc.nasa.gov/about\_swift/xrt\_desc.html}, for example.

Other software could also be developed to robustly detect very short duration outbursts that may not be identified using the current variability tests enabled in the pipeline, or which may have too few counts to be detected as a source in a long observation, because of the background contribution. Examples of such objects are distant short gamma-ray bursts ($<$2 s) which would have much of their redshifted emission in the X-ray domain, the electromagnetic counterparts of gravitational wave events, type-I X-ray bursts from neutron stars in other galaxies, or possibly X-ray counterparts to fast radio bursts. Again the PI could be alerted of these events in their data. Having this information shortly after the observation may also allow the follow-up of gamma-ray burst afterglows, identify if a gravitational wave event was observed serendipitously in the {\it XMM-Newton} field of view or give clues to the nature of the as yet unknown nature of the fast radio bursts \citep[e.g.][]{hess18}.

\section{Concluding remarks}\label{sec5}

Identifying more TDEs with {\it XMM-Newton} will not only help to increase the number of known IMBH, it will also be one way to help understand the relationship between the X-ray and optical/UV emission as the OM points in the same direction as the X-ray telescopes, allowing many TDEs to have contemporary X-ray and  optical/UV observations. In addition, studying the X-ray spectra of TDEs when they are bright will allow us to probe the physical mechanism behind super-Eddington accretion, which may have played a significant role in the growth of SMBH. 

Whilst IMBH in orbit with another compact object could instead be detected with gravitational waves, current facilities can not detect black holes with masses in excess of a few hundred solar masses \citep[e.g.][]{aasi14} and therefore most of the IMBH mass range can not be exploited. Only future facilities, such as {\it LISA} will be able to detect IMBH \citep{bara15}, but these observations will not take place for more than fifteen years. It is therefore timely to exploit X-ray observations today to maximise our understanding of IMBHs and the formation and evolution of massive black holes. It is clear that if {\it XMM-Newton} were to function as suggested here, it would also be beneficial for finding other highly variable X-ray emitting objects and allowing their rapid follow up.







\bibliography{WebbV2}%



\end{document}